# Usage Bibliometrics as a Tool to Measure Research Activity

Edwin A. Henneken, Michael J. Kurtz
Smithsonian Astrophysical Observatory, 60 Garden Street, Cambridge, MA 02138, USA

## Abstract

Measures for research activity and impact have become an integral ingredient in the assessment of a wide range of entities (individual researchers, organizations, instruments, regions, disciplines). Traditional bibliometric indicators, like publication and citation based indicators, provide an essential part of this picture, but cannot describe the complete picture. Since reading scholarly publications is an essential part of the research life cycle, it is only natural to introduce measures for this activity in attempts to quantify the efficiency, productivity and impact of an entity. Citations and reads are significantly different signals, so taken together, they provide a more complete picture of research activity. Most scholarly publications are now accessed online, making the study of reads and their patterns possible. Click-stream logs allow us to follow information access by the entire research community, real-time. Publication and citation datasets just reflect activity by authors. In addition, download statistics, derived from these click-streams, will help us identify publications with significant impact, but which do not attract many citations. Click-stream signals are arguably more complex than, say, citation signals. For one, they are a superposition of different classes of readers. Systematic downloads by crawlers also contaminate the signal, as does random browsing behavior. We will discuss the complexities associated with clickstream data and how, with proper filtering, statistically significant relations and conclusions can be inferred from download statistics. We will describe how download statistics can be used to describe research activity at different levels of aggregation, ranging from organizations to countries. These statistics show a strong correlation with socio-economic indicators, like the GDP. A comparison will be made with traditional bibliometric indicators. Since we will be using click-stream data from the Astrophysics Data System (ADS), we will argue that astronomy is representative of more general trends.

# Introduction

The standard indicators to measure the quality and quantity of scholarly research are funds expended, number of papers published, and number of citations to those papers and measures derived from these citations. Since the turn of the century a fourth key indicator for research assessment has arisen: measures of the use of the (now almost exclusively) digital research documents. The concept of "research documents" represents a more general class of expressions of research than "scholarly publications", because it contains anything that may get published during research. This study is confined to scholarly publications, that is articles in scholarly journals. Scholarly research can be represented by identifying a "research cycle"; research consists of activity expended in the various stages of this cycle (illustrated in figure 1). We assume that taking the publication stage of the research cycle as a proxy will sufficiently represent "research activity". Also, from a practical point of view, the publication stage is the only stage in this cycle that allows for clearly quantifiable metrics. We therefore focus on usage of scholarly publications.

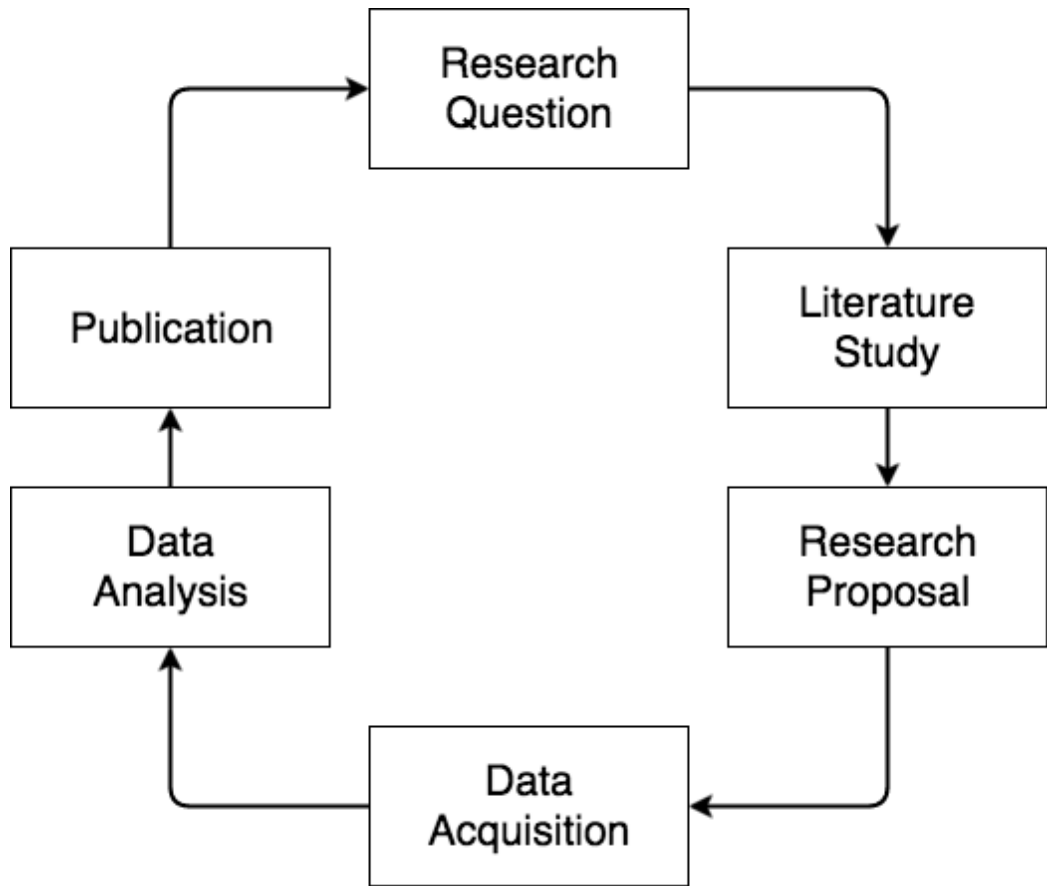

*Figure 1: graphical representation of research cycle*

The use of this usage information is in its infancy. The leading assessments of the quality and quantity of research on a country basis (Science & Engineering Indicator [1], for a given year), a university basis (Times Higher Education [2] or ARWU [3] ranking, for a given year), or on a journal basis (Impact Factor [4], Eigenfactor [5,6] or Source Normalized Impact per Paper [7], for a given year) do not use usage information.  The only widespread use of digital download records is by librarians making purchase decisions, aided by the COUNTER [8] standard for a given year, continuing their practice from the print era.

Since the first obsolescence function based on digital downloads was published 20 years ago [9] there has been an avalanche of work on the nature of digital download information. The review by Kurtz and Bollen [10] contains 171 references, more recent work with extensive discussions and bibliographies include [11, 12, 13, 14, 15, 16, 17, 18].  Download information is now commonly found on article pages at journal web sites and on various aggregator web sites. Perhaps the most influential of these are the article download counts for a given year by the SSRN.

Before usage information can be widely used in making practical decisions some obvious "problems" need to be addressed. Just as the most popular wine, novel, politician is not necessarily the best one, pure usage counts of scholarly article downloads may not be an accurate measure of research activity.

Users of scholarly research articles can be crudely divided into four categories [10, 11]: researchers, practitioners, students, and the general public. It is not uncommon that researchers represent just a small fraction of the total use. The number of health care practitioners dwarfs the number of medical researchers, for example. Students [19] use the scholarly literature very differently from researchers. In astronomy the number of interested lay people exceeds the number of research astronomers by a factor of perhaps 10,000; the number of serious amateur astronomers is about 100 times the number of researchers, while the number of citizen scientists on a single astronomy project, Galaxy Zoo, is a factor of 10 larger than the world total of researchers in astronomy [20]. All these groups can and do access and read the scholarly literature. However, just the users class of researchers contribute to usage corresponding to research activity.

In this paper we use download measures to assess the quantity and to some degree quality of astronomy research, with the expectation that the techniques will also prove useful in other fields. We are aided in this by the fact that there are few professional astronomers who are not researchers; there is no such thing as applied astronomy. This simplifies the task compared with a field such as medicine, where a majority of professionals are not researchers. There are two different aspects to download measures: who is downloading, and what is downloaded. Both are quantifiable signals and we will use both of them in our analysis. In order to get meaningful results, measures for both aspects are essential.

To some degree, research activity is related to economic trends. In previous publications we have explored relationships between economic indicators and research-related indicators [21, 22, 23, 24]. This present paper substantially expands on the download analysis in these publications and shows that the data used in this analysis are consistent with the conclusions from these earlier studies.

# Definition of Terminology

Use and usage are terms that seem to have a range of different meanings in the literature on bibliometrics. Other terms often encountered are "hits", "clicks", "reads" and "downloads". Used as a measure, it reflects aspects of user interaction with a digital service, in this case the Smithsonian/NASA Astrophysics Data System (ADS). During a session, users typically access different types of information by clicking on the appropriate links. In this study, we restrict ourselves to "downloads", which is defined as clicking on a link that will take a user to the full text version of an article (either stored locally on an ADS server or stored externally). In the presentation and discussion of our results, we will use the terms "downloads" and "usage", but they both will refer to the act of getting to the full article text.

The terms listed in table 1 will be used frequently in the description of data (among others) and are particularly important with respect to creating and interpreting data. Hence, it's important to define them.

| term | definition |
|------|-----------|
| entity | A group of people generating publications, which can be identified through queries in the ADS, and whose interaction with the ADS can be easily identified in the usage logs. The following examples come to mind: country, institute/organization |
| frequent user | This is an ADS user identified from the usage logs as one with at least 100, but no more than 1000 downloads per year. The reason for the upper limit is the fact that there are download sessions that cannot be associated with one particular individual (like computers in libraries). |
| Main astronomy journals | The Astrophysical Journal, The Astrophysical Journal Letters, The Astrophysical Journal Supplement Series, The Astronomical Journal, Monthly Notices of the Royal Astronomical Society, Astronomy & Astrophysics |

*Table 1: Definition of frequently used terms*

## Data

The data in this study have been derived from the usage logs of the ADS. These usage logs contain the user interactions with the user interface, recording the nature of the interaction (the type of information being requested), the time of this interaction (the date and time in EST), the identifier associated with the user and the IP address of the source requesting the information. The originating country of a request is derived from the hostname of the request, which in its turn is derived from the IP address. In the mapping from hostname to country, we associated the top level domains ".net", ".edu", ".gov" and ".mil" with the USA.

When logging user interactions with the ADS, we attempt to filter out robots based on our knowledge of so-called "User Agents" and origin IP addresses of these requests. Since these robot requests are mostly for metadata, missed robot requests in our filtering will not contaminate our data, because we are focussing our study on downloads. Figure 1 shows the number of users in various categories over the period of our analysis (2005-2015). The line representing the "remainder" in this diagram represents situations like computers in libraries. In an earlier study [15] we observed that the median of the number of reads for frequent users is fairly constant at a value of about 21 reads per month. Since this number includes more than downloads, we decided to take 100 downloads per year as lower boundary for the frequency interval associated with "frequent users". Usage data for frequent users suggests a read to download ratio of between 2 and 3. A second argument indicating that our choice of the definition for frequent users is meaningful, is illustrated by the number of frequent users when restricted to downloads of publications from the main astronomy journals (line with open triangles in figure 2). This restriction will underestimate the number of research astronomers, so it is a lower limit. In the period of 2005-2015, the total number of members of the International Astronomical Union grows from around 9,000 in 2005 to around 12,000 in 2015, represented by the dashed line. We only have actual IAU membership numbers for the period of 2008-2015. The fact that this line is bounded by the two lines representing both types of frequent users is a strong indication that our definition of frequent user can be interpreted as a representative definition.

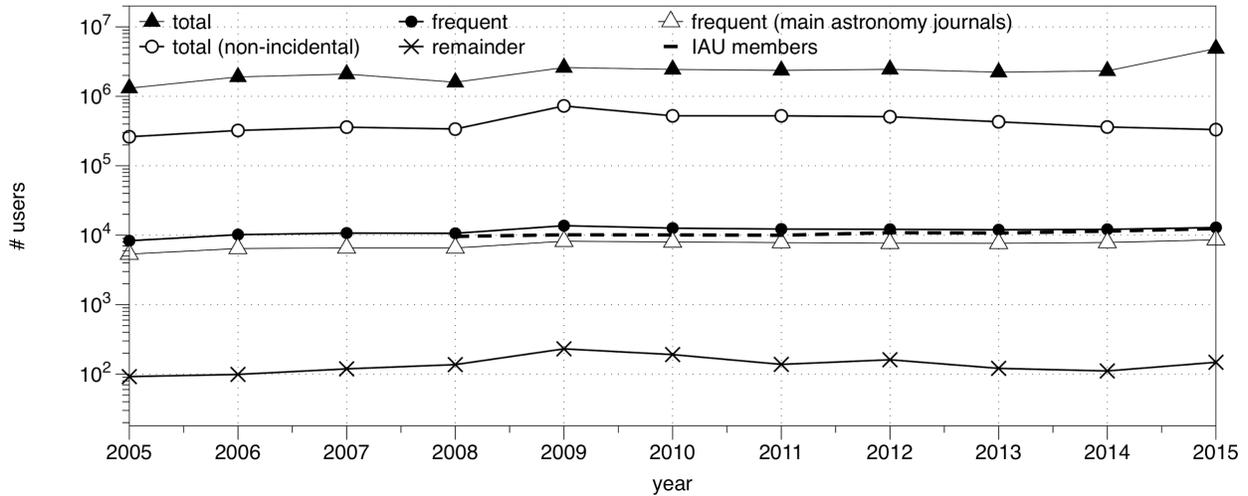

*Figure 2: The number of ADS users during the period of analysis, based on the number of yearly downloads. The line with solid triangles show the total number of users for each year, the line with open circles shows the total number of users who downloaded at least 1 publication (this excludes those users who just look at an abstract), the line with solid circles show the number of users who accessed full text between 100 and 1000 publications per year and the line with crosses represents the remainder. The line with open triangles shows the number of users who accessed the full text for the main astronomy journals between 100 and 1000 publications per year. The dashed line represents the total number of IAU members.*

Figure 1 also characterizes the type of signal we are looking at in this study. We will be looking at the access of full text ("downloads") by frequent users, which is roughly two orders of magnitude smaller than the total access. Also, as shown in an earlier study (figure 3 in [24]), the readership pattern of frequent users is significantly different from that by incidental users. Figure 2 indicates that the class of frequent users of the ADS exceeds the number of professional astronomers, the excess mostly likely consisting of physicists and engineers. Since most professional astronomers are using the ADS on a daily basis, it makes sense to focus on the field of astronomy in our analysis, because this will result in signals that can be regarded as truly representative for the entire field.

Why do we focus on the downloads by frequent users? Essentially because all authors are ADS users, but not all ADS users are authors. A comparison of the obsolescence functions of reads and citations makes this abundantly clear. Figure 3 (taken from [24]) illustrates this fact. From the ADS usage logs for January and February of 2008, we display the usage by frequent users of the ADS with those coming in from Google Scholar (representing incidental users) and compare these signals with citation rates and total citations. This analysis has been restricted to publications from the main astronomy journals. The fact that the relative use for frequent users follows the citation rate, rather than total citations, as function of publication year illustrates why we should focus on usage by frequent users. Citations, by their very nature, are deliberate and by using usage by frequent users, we expect to distill a signal that is expected to represent actions by authors with the least amount of noise. Note that an additional interpretation of figure 3 is that Google Scholar does not provide results that researchers are looking for.

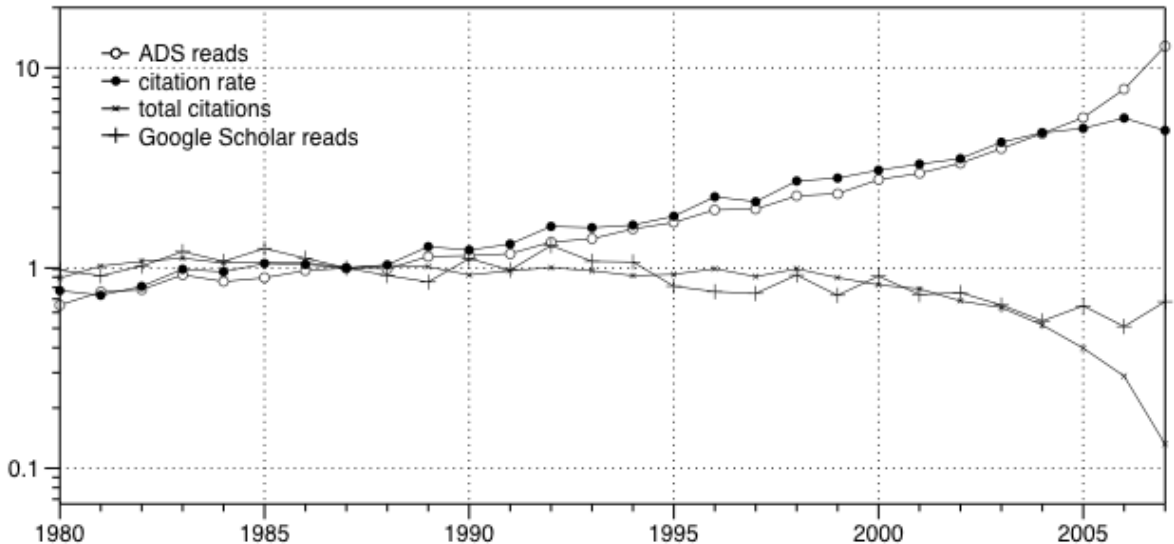

*Figure 3: Comparison of readership patterns from ADS and Google Scholar queries, as observed in the ADS access logs. The line marked with open circles shows the readership use by people using the ADS search engine. The line marked with '+' corresponds with the readership use by people who used the Google Scholar engine. The line marked with solid circles shows the citation rate to the articles, while the line marked with 'x' represents their total number of citations.*

The other component of research activity considered in our study consists of the scholarly publications generated by entities analyzed. For each year in the period analyzed, a bibliography is compiled for any given entity using the query capabilities of the beta-version of the ADS, code-named "Bumblebee" (http://ui.adsabs.harvard.edu). The main reason for using this version of the ADS, rather than the older version, ADS "Classic", is the fact that it supports a rich query language, allowing queries that involve affiliation information. With a couple of exceptions, the bibliographies were determined with the following query

    aff:"<affiliation string>" year:2005-2015

We used the API of ADS "Bumblebee" (https://github.com/adsabs/adsabs-dev-api) to generate the bibliographies with metadata identifying the publication, the affiliation of the authors and the refereed status of the publication, indicating whether it is a refereed or non-refereed publication. The few exceptions to this approach are those for which the ADS already contains a curated bibliography in its database, in which case the query just needed to retrieve those entries from this bibliography for the year range considered. Table 2 shows an overview of publication number for the start and end of the period, for a number of entities (with a wide range in size)

| entity | total | | refereed | | main astronomy | |
|---|---|---|---|---|---|---|
| | 2005 | 2015 | 2005 | 2015 | 2005 | 2015 |
| Yale | 677 | 1130 | 502 | 945 | 79 | 194 |
| Princeton | 1479 | 2338 | 1157 | 1603 | 195 | 264 |
| CfA | 2541 | 1950 | 713 | 1061 | 497 | 824 |
| NOAO | 283 | 251 | 232 | 246 | 218 | 238 |
| Canada | 5785 | 9055 | 4209 | 6494 | 395 | 893 |
| The Netherlands | 3444 | 6120 | 2697 | 4169 | 442 | 841 |
| Argentina | 899 | 1376 | 793 | 1062 | 85 | 139 |
| India | 3484 | 9278 | 3190 | 6927 | 153 | 339 |

*Table 2: Overview of publication numbers for the start and end of the period, for a number of entities*

We will work with a data set with maximum homogeneity, by restricting both usage data and publication data to the main astronomy journals. This way we can be assured that all necessary metadata will be available for article selection.

# Usage and Research Activity

One of the "products" of research activity is the generation of scholarly publications. It seems reasonable to assume that, as part of the preparation process, publications are read and may get added to the bibliographies of future publications. Since the ADS is the main literature discovery tool in astronomy, we assume that the publications read during this preparation process are found using the ADS. This model implies a number of questions:

1. Is there a correlation between the publications generated by an entity and the ADS usage by people associated with that entity?
2. Assuming that at least a fraction of downloaded papers will get cited in the publications generated, is there enough signal to detect that relationship?

In addition to the question whether there is a correlation between downloads and number of publications, we also should look into the similarity of both sets. If our assumption is true that people download publications that serve as foundation for their publications, you would expect an overlap between the two sets. This leads to a third question:

3. How similar are the set of publications cited by the publications generated by an entity and the publications downloaded by people associated with that entity?

Before exploring these questions, we need to address the relationship between the actors in the usage and research activities. In the section on the data, we looked at usage data and publication data as separate quantities, but since the ADS lies at the center of both, we expect some common sense correlations. For the authors, associated with a given entity, especially when they are the first author, we expect them to be among the frequent ADS users for that entity. This should be particularly true when we restrict ourselves to publications downloaded by the frequent ADS users associated with that entity. For example, given all the papers published in one of the main astronomy journals in 2005, where one of the authors is affiliated with an institute in Belgium, what is the correlation between number of authors who appeared as first author in this set and the number of frequent users from Belgium in 2005? Figure 4 shows this relationship for a number of countries and institutes.

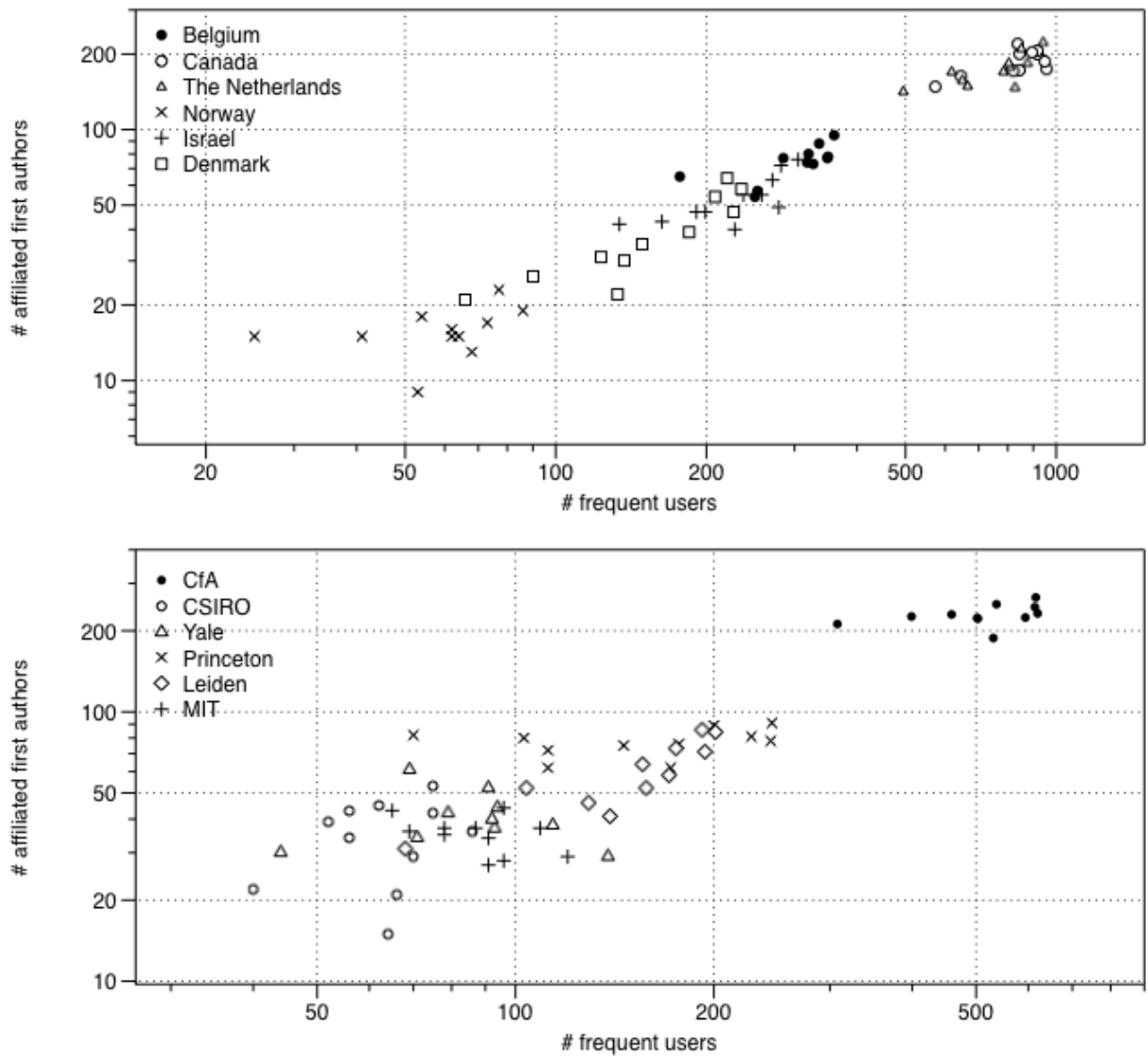

Figure 4: Relationship between the number of frequent users and the number of affiliated first authors for a number of entities. Every data point corresponds with one year for the entity displayed. Top panel: countries. Lower panel: institutes

For countries, we also expect a significant correlation between the number of affiliated first authors and the number of national members of the IAU. This is shown in figure 5.

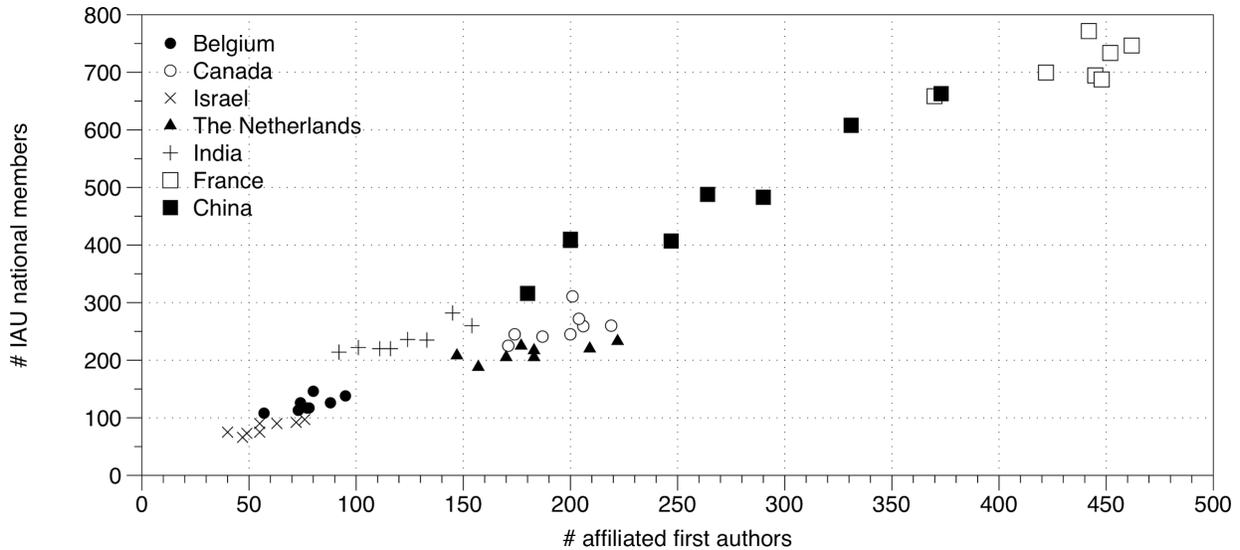

*Figure 5: Relationship between the number of affiliated first authors for a number of countries and the number of national members of the IAU. Every point represents a year in the period 2008-2015 (no IAU data was available prior to 2008)*

Figures 4 and 5 show that our data is consistent with common sense expectations: there is a significant overlap between the population downloading publications on a regular basis and publishing articles in main astronomy journals.

Next, we will explore the relationship between downloads and generated publications for a range of entities. As before, we will focus our analysis on frequent users and publications from the main astronomy journals. Figure 6 shows the numbers of downloads by frequent users of main astronomy papers recently published versus the number of main astronomy papers where one of the authors is affiliated with the entity under consideration.

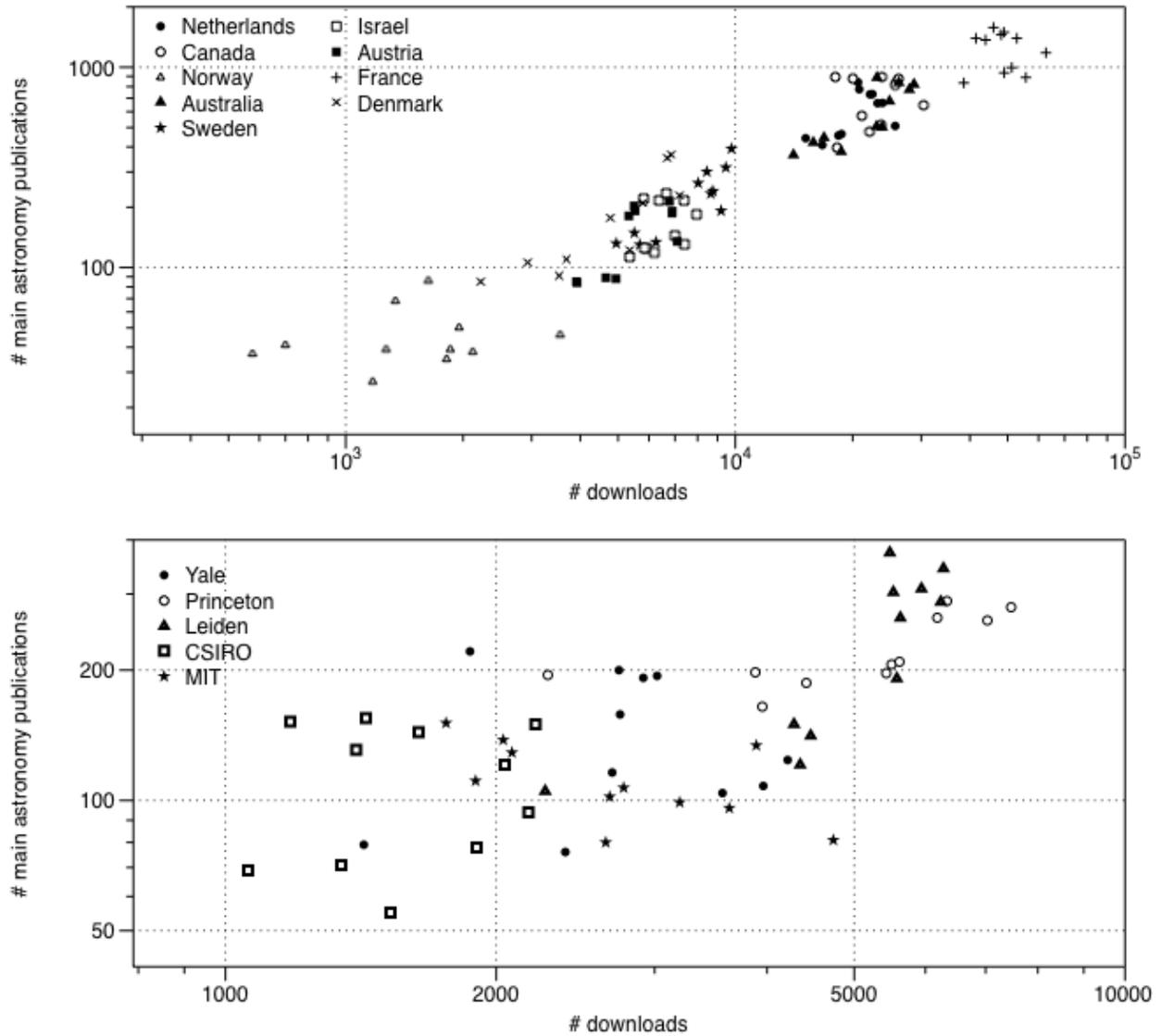

*Figure 6: Number of downloads by frequent users of main astronomy papers recently published versus the number of main astronomy papers where one of the authors is affiliated with the entity under consideration. Every point represents a year in the period 2005-2015. Upper panel: relationship for countries. Lower panel: relationship for institutes*

The correlation between download and publication numbers has a scalar character. It does not provide any insight in the similarity of the downloaded and published articles. This is an aspect we can explore in several ways. The following quantities will be used to explore the similarity between downloaded publications and those cited, in a given year $Y$ for a specific entity $E$:

| $R^Y(E)$ | Publications in main astronomy journals downloaded by frequent users associated with E |
|---|---|
| $P^Y(E)$ | Publications in main astronomy journals where the first author is affiliated with E |

| $C^Y(E)$ | All publications in main astronomy journals cited by the publications in $P^Y(E)$ |
|---|---|

Every publication has a publication year. Figure 7 shows a number of relationships based on publication year. One signal we have observed is, for a given year and entity, the set of downloaded publications in the main astronomy journals, by frequent users, with a publication year in the interval starting in 1980 and ending in the year under consideration. For this signal we derive the following two quantities: for each year in the range of publications years, the number of downloaded publications, normalized by the total number of publications in that set. The other quantity is the number of unique publications with a given publication year, normalized by the total number of publications in the main astronomy journals for that year. We derive the same quantities for another signal: all papers from the main astronomy journals and a publication year in the same range, cited by the publications in the bibliography for the entity in the year under consideration. Figure 7 shows the results for The Netherlands (E) in 2015 (Y).

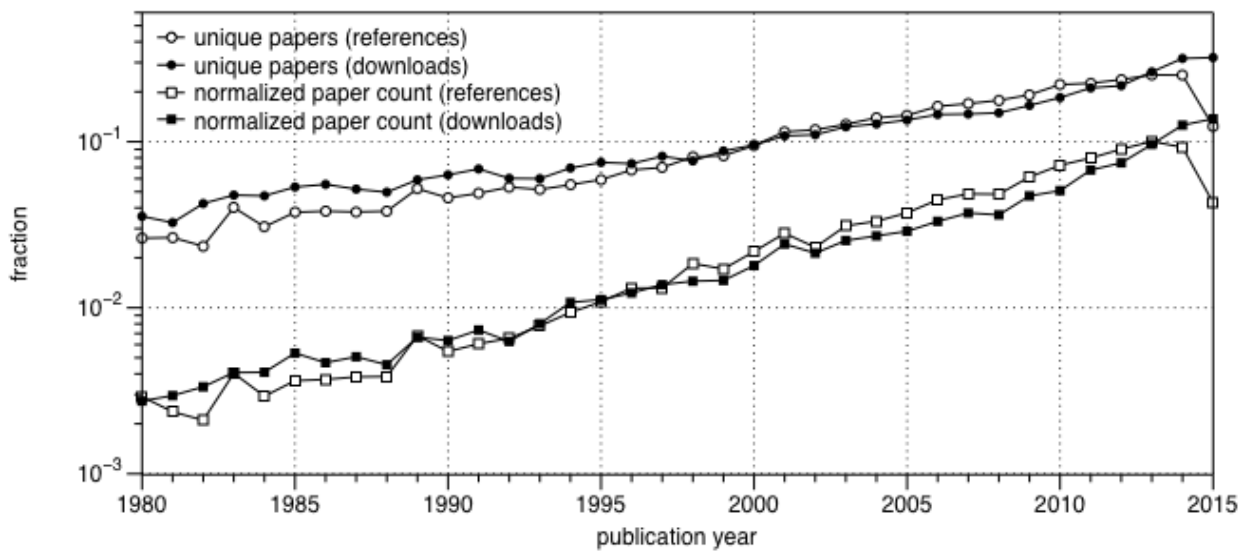

*Figure 7: For The Netherlands ("entity") and the year 2015, this figure compares the distribution of publication years, in the range of 1980 through 2015, in the lists of downloaded main astronomy publications by frequent users and those cited by main astronomy publications in that year. The lines with open and closed circles show the fraction of unique publications in the citation (open) and download (closed) lists. The lines with squares show the normalized publication numbers in the citation (open) and download (closed) lists.*

Finally, we consider similarity on the most granular level: using individual publications as the data in the comparison. Assuming that researchers, affiliated with an entity, actually read the publications (at least to a non-negligible degree) they cite in their scholarly papers, you expect an overlap in those citations and the publications downloaded by frequent users from that same entity. The critical assumption here is that the ADS is used to get to the full text (either hosted locally on the ADS servers or external to the ADS). For both set of publications, we have lists of identifiers, uniquely associated with publications. We will use these identifiers in our analysis. For a given entity in a given year, the main astronomy papers cited in the first-authored papers are compared with the downloaded main astronomy papers by frequent users, associated with the same entity. We will consider the similarity measured by the fraction of overlap between the two sets. Figure 8 shows results for a number of countries.

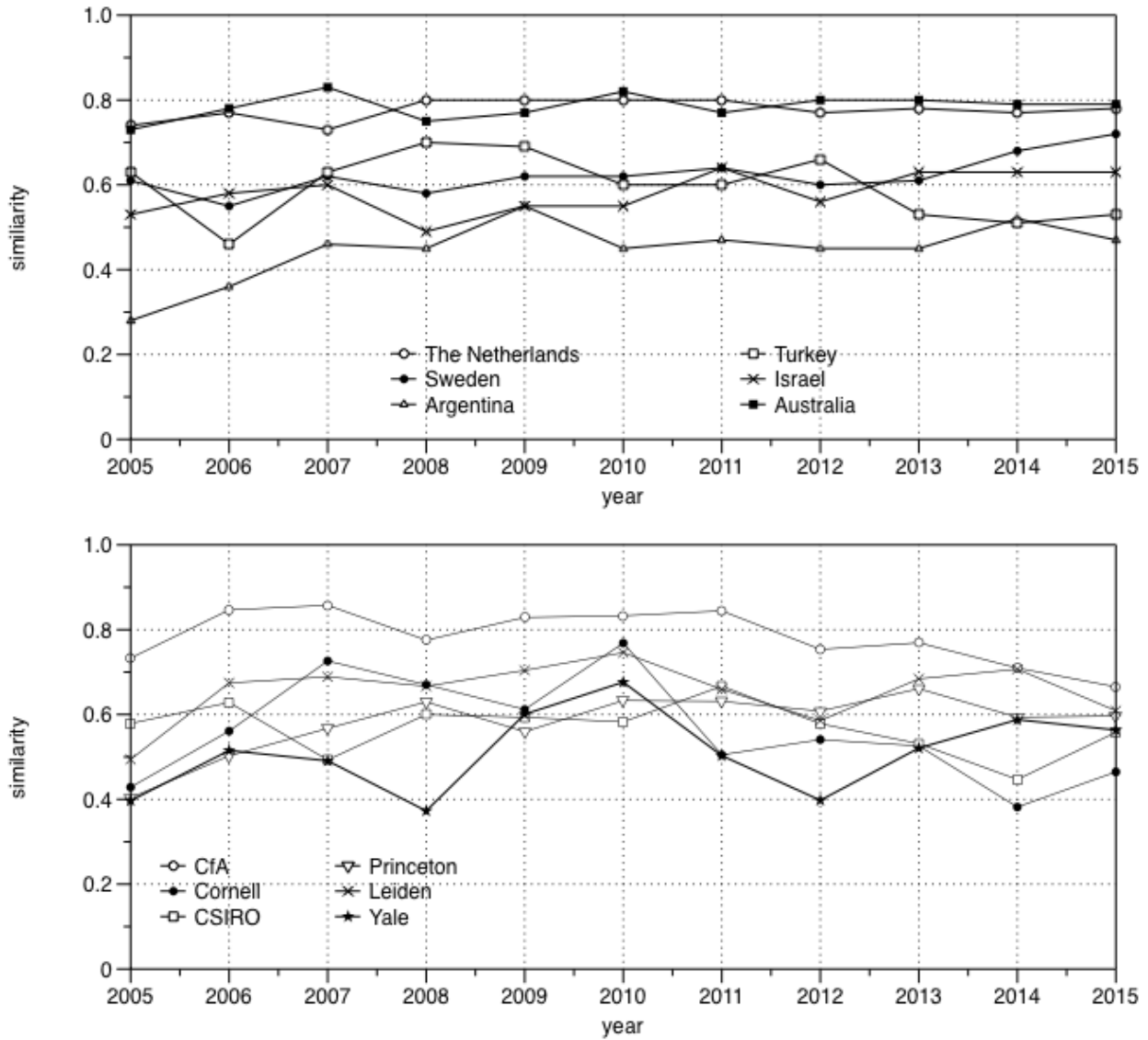

*Figure 8: The fraction of overlap between the sets of main astronomy papers cited in the first-authored papers are compared and the downloaded main astronomy papers by frequent users, associated with the same entity. Upper panel: relationship for countries. Lower panel: relationship for institutes*

# Traditional Indicators

Bibliometrics is an example of a discipline that provides tools for quantifying research output and its impact. As mentioned before, we assume that research output, in the form of publications, can be regarded as a proxy for research activity. In this context, bibliometrics provides a suite of measures that can be seen as research metrics. These bibliometric measures are used to quantify a degree of research output. Traditionally, citations are used as fundamental building blocks for these measures. There are many citation-based research productivity measures, ranging from straightforward ones, like total citations, to complex indicators, like the Tori indicator [16]. The one thing they have in common is that, in fact, they are a measure for impact, rather than a direct measure for research activity. Of course, without research activity, there is nothing to cite, so in this sense they are an indirect measure for research activity. Even when e-prints from a repository like arXiv.org are used in conjunction with these measures, they do not provide the immediate sense research activity that usage-based measures provide. By comparing citation-based measures with those based on usage, you look at more than just research activity. This comparison also involves research quality.

With the correlation found between usage (in the form of downloads by frequent users) and publications, you would expect significant correlations with citation-based indicators. The act of publishing generates citations. In figure 9 we explore the relationship between the number of publications (in the main astronomy journals) generated for an entity in a year and the value of the h-index, for these publications, in the next year.

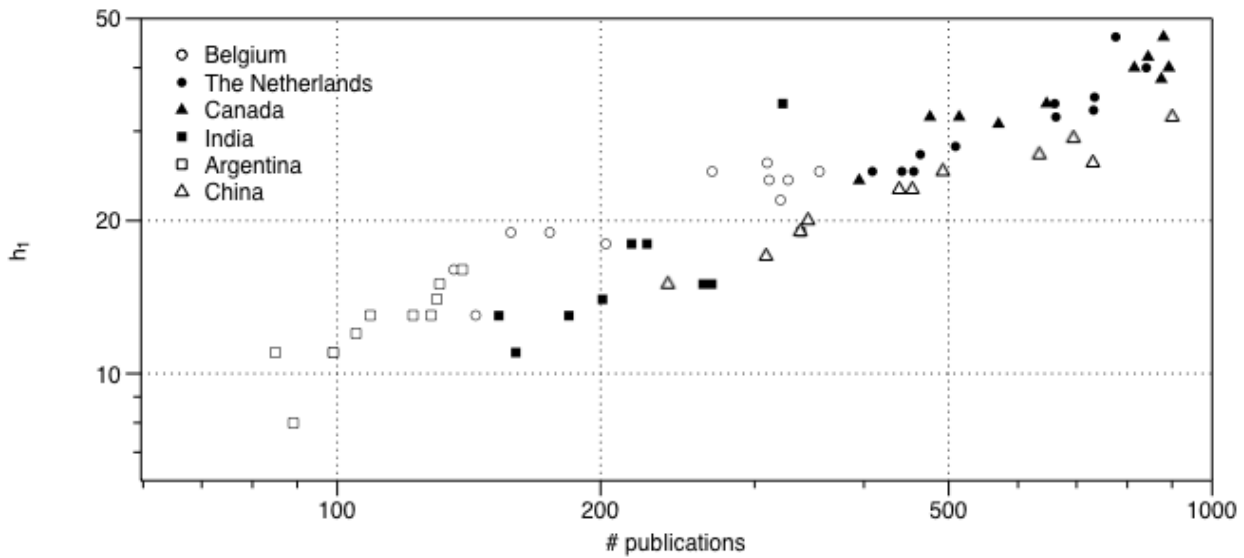

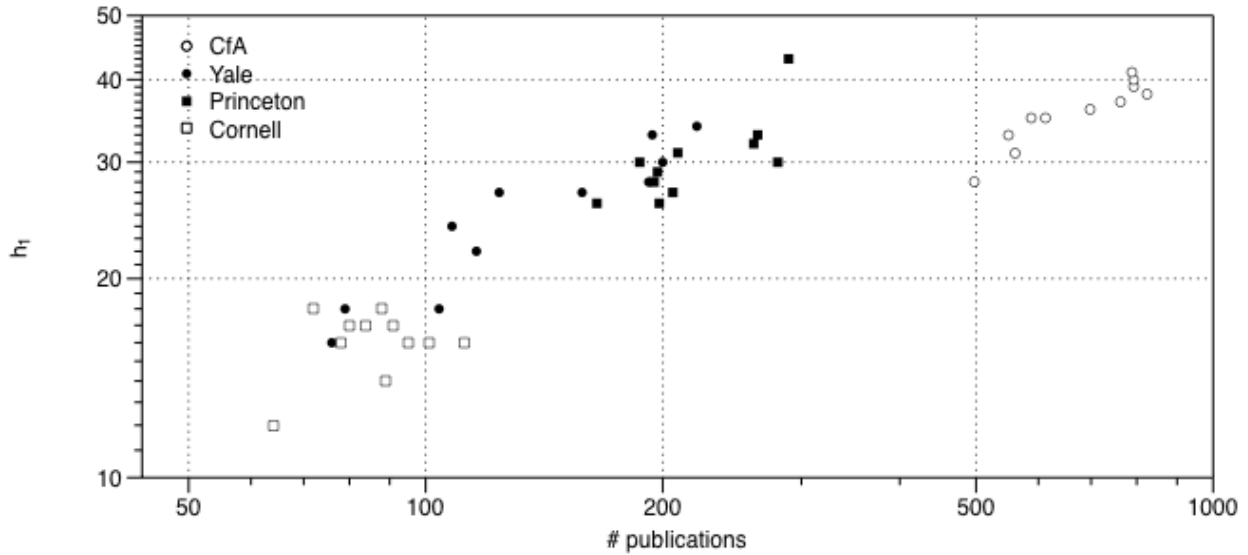

*Figure 9: The number of publications in the main astronomy journals associated with a entity compared to the h-index for these publication in the next year. Top panel: countries. Lower panel: institutes*

# Discussion

Our selection of data for the analysis put forward in this publication has been guided by the following observation: while citation is a deliberate, public act, usage is private act, with a wide range of types. Citations are solely generated by authors, while usage is not solely the result of actions by authors. Taking publications as a measure for research activity, we necessarily need to focus on usage patterns associated with ADS users who are most likely to be authors. Because we have unique identifiers associated with users, we are able to do this. Figure 2 illustrates that, with a proper definition of usage frequency, a class of users can be delineated that strongly correlates with the size of professional researchers. With lower and upper boundaries set to 100 and 1000 downloads per year, over the period covered by this paper (2005-2015), the number of frequent users closely resembles the number of IAU members for the period with available data (2008-2015). For a number of entities, we determined the set of publications in the main astronomy journals where one of the authors has an affiliation associated with that entity. From these sets, we created subsets where the first author is affiliated with a particular entity. For each year in our period of analysis, we compared the number of first, affiliated authors with the number of frequent users for that entity. This comparison shows a very strong correlation for countries and strong correlation for institutes, but definitely more spread. Examples of sources that contribute to spread are the fact that not all frequent users are authors and time lapses inherent to the act of publishing (particularly relevant for publications early and late in each year). In countries with multiple institutes, especially those with higher numbers of frequent users and authors, spread is less because of an averaging effect. In the case of countries, figure 5 compares the number of affiliated first authors with the number of national IAU members, for the period 2008-2015. Affiliated first authors can be visiting scholars from another country or simply not be a member of the IAU. Also, not all national IAU members are active authors. Nevertheless, figure 5 shows a strong correlation between both numbers. Taken together, figures 2, 4 and 5 make a strong case for being able to trust that the class of frequent users, derived from the ADS usage logs, sufficiently represents those scholars that generate research activity in the form of publications, for a given entity.

In the next stage, we compare the signals, at various levels of granularity, generated by the two populations we have defined: authors affiliated with an entity and frequent users associated with that entity. The author-centric data consists of publications generated in a particular year (and the publications cited in their respective bibliographies) and the data for the frequent users consists of lists of publications downloaded by them in a particular year. In both cases, publications are identified by a unique entity. At the least granular level, we compare just numbers, specifically the number of publications from the main astronomy journals, downloaded by a frequent user associated with an entity, and the number of main astronomy publications where one of the authors is affiliated with that entity. Figure 6 shows that the correlation is the strongest for countries. As noted before, there are the effects of missed data on the start and end of each year, but the effect of this gets dampened on the scale of a country. The smaller amounts of data on the level of institutes results in a larger amount of scatter. At best, the results in figure 6 show that there is a relationship in the mean. By comparing scalar quantities we have removed any information on what actually is downloaded and published.

Ideally, authors should read every publication that they intend to cite in the paper they are writing. However, this is unlikely to be true [26]. The question is whether this is true enough. Assuming that the populations of frequent users sufficiently represent authors, we still expect a lot of scatter on the detailed level of actual downloaded publications. Publications are downloaded that for various reasons do not get cited or get cited in a paper which appears in the next year. Figure 7 explores relationships at a more granular level, but not yet at the most granular level of using individual publications in comparisons. Figure 7 was constructed by using one particular aspect of article metadata: the publication year. With the publication year we examine the similarity in obsolescence functions for cited and downloaded publications. We explored these quantities for one particular entity and year, but the results found hold true for a wide range of either. The similarity of the obsolescence functions indicates that the correlation found earlier, in the least granular level, is still present at the more granular level of using the publication years of individual articles.

Before going to the level of individual publications, there is at least one additional approach for comparing the results for both populations. This approach would give an estimate for the similarity in subject matter between the published publications and those downloaded by frequent users, for example by using a clustering algorithm (like k-means or Louvain clustering based on keywords). We will not consider this approach in this publication, but it may be a subject for a future publication.

Finally, we explore the similarity between downloaded publications and the cited papers, for a range of entities. We selected the most straightforward comparison: calculate the fraction of overlap between both sets. Figure 8 shows the results for a number of entities. Would these fractions have been significantly smaller if the downloads had been a random selection of publications from the main astronomy journals? Figure 10 shows results for a number of entities. Every "random" value is the average of 10 samples.

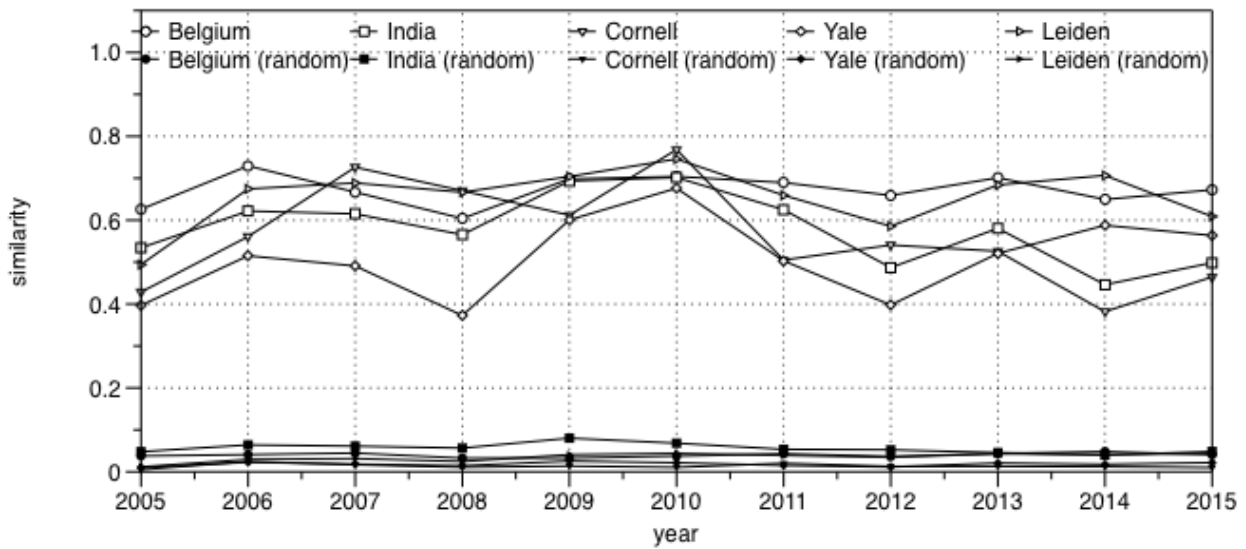

*Figure 10: The fraction of overlap between the sets of main astronomy papers cited in the first-authored papers are compared and the downloaded main astronomy papers by frequent users, associated with the same entity. Lines with open symbols show the actual data and lines with corresponding solid symbols show the results if the downloads had been random.*

For both countries and institutes we find moderate, yet significant similarities between main astronomy publications, downloaded by frequent users and the main astronomy publications cited in the publications for a given entity and year. We wouldn't expect this similarity to be more than modest because of the reasons we mentioned earlier.

Is there a relation with traditional bibliometric measures? It is difficult to use traditional bibliometric measures to quantify research activity, in particular if you are interested in a measure that is very close, temporally, to when the research is actually happening. For any measure based on citations, the articles will have to have been published and available long enough to accumulate a reasonable amount of citations. By considering citations as a measure, you are immediately attaching a quality assessment to your analysis. A usage-based measure, like the READ10 indicator [22], based on current reads of publications, is an alternative, but it is still an impact measure. This does not mean that there is nothing to compare. Given the fact that most of these traditional measures are citation-based, any relationship is indirect, circumstantial at best. We have shown that download activity correlates with publication numbers. Publications result in citations. Taking into account that there is a lag for citations to accumulate, there may be a correlation between current downloads and a citation-based indicator "later" (say, the following year). The results in figure 9 show this to be a correct assumption. Kurtz et al. [22] showed the relation between downloads of an author's work and citations to that work, both on an individual author basis, and summed over research institutes.

Eichhorn et al. [21] first showed the relation between the number of times individuals in a country download and astronomy article and the GDP of that country, and in an earlier publication we showed a relationship between general usage and the Gross Domestic Product (GDP) per capita for a range of geographic regions [24]. We found that growth in GDP per capita in general translates in an increase in ADS usage. Is there a similar trend for just downloads by frequent users? Figure 11 indicates that, with a few exceptions, an increase in GDP per capita results in an increase of downloads by frequent users, implying increased research activity. An exception is China, where it seems that research activity does not keep up with the explosive economic growth.

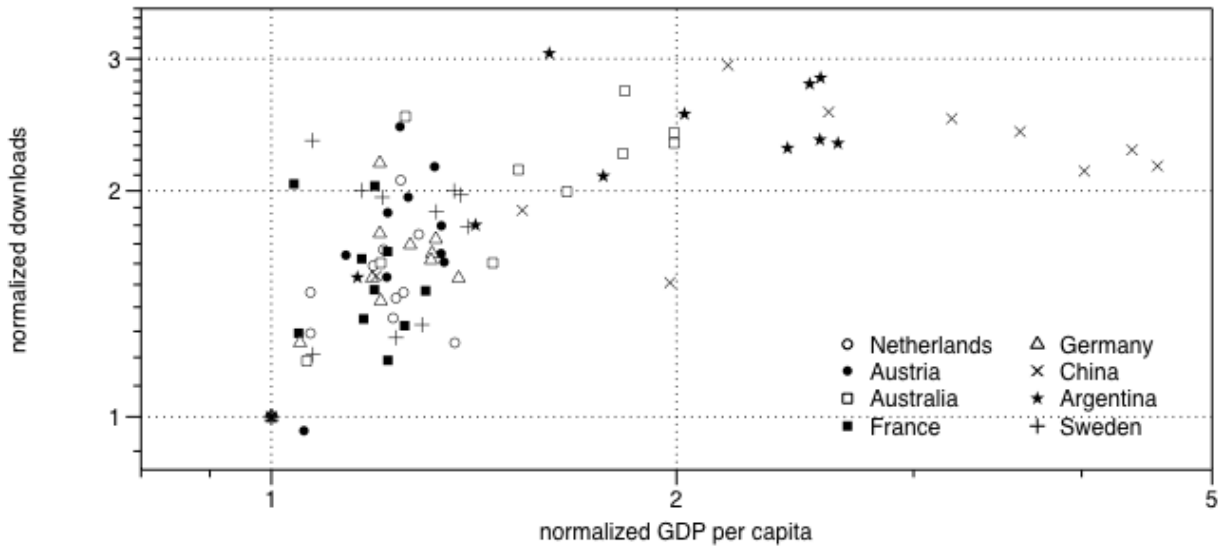

*Figure 11: GDP per capita (current USD) versus number of downloads by frequent users. Both quantities have been normalized by their values in 2005. Every point corresponds with a year in the range 2005-2015.*

Eichhorn et al. [21] first showed the relation between the number of times individuals in a country download and astronomy article and the GDP of that country. Kurtz et al. [23] showed that the number of downloads in a country is proportional to that country's GDP squared divided by it population. In that publication, they also showed that that the number of downloads by researchers of astronomy articles, better predicts the mean of the number of articles published and the citations to those articles. All are measured as fractions of the world total.

# Concluding Remarks

When a researcher cites a publication in an article, it is a public, deliberate act. The only room for interpretation is the sentiment of the citation. From a process point of view, because of this deliberate nature, there is no noise component in the citation signal. This cannot be said for the process of usage. For any entity where some level of scholarly research is performed, literature discovery is an essential ingredient in the research lifecycle. However, this is a process with many components, depending on the goal of the literature search. All of this activity, for all entities on all levels, comes together for a service provider like the ADS. Somewhere, buried in the millions of yearly interactions with the service, are the signals that represent the act of gathering the necessary literature for producing scholarly articles. Although "research activity" consists of many components that do not involve literature search and writing, we feel that the act of generating scholarly articles is a measure that is a good proxy for research activity. Key to being able to find these signals is the ability to associate sessions with individuals. It is not sufficient to use IP addresses to identify sessions. For one, researchers are mobile and some institutes proxy their requests through one single host. We noted the presence of library computers in figure 2 and how these result in usage of multiple users ending up being indistinguishable. For usage information to have any use, either for bibliometric analysis or to enhance the discovery experience of a service, a lot of meaningless systematic (e.g. robots) and random (e.g. incidental users entering through Google) signals need to be removed. We showed the dramatic difference in orders of magnitude between incidental use and the use by people who use the ADS professionally. When thinking about a very specific class of researchers, namely authors, it is important that all authors use the ADS, but that not all users (even frequent users) are authors. Going back to our original question of quantifying research activity, this emphasizes the need for being able to properly identify these frequent users. In this study we showed that for the ADS we can identify a class of frequent users and that there is convincing evidence that these users represent the population of active researchers (and even authors) in astronomy. Having identified these frequent users for various entities and having constructed bibliographies for these entities for the time period of 2005-2015, we have shown that there is a significant correlation between the number of frequent users and first, affiliated authors for a range of entities. For countries we also showed a correlation between the number of first, affiliated authors and the number of national IAU members. Based on this evidence we argue that the class of frequent users represents the authors making use of the ADS, while in the process of writing scholarly articles. For the main astronomy journals, the download activity by these frequent users correlates with publications on multiple levels of granularity. This correlation is stronger for countries than for institutes. Our conclusion is that download activity for main astronomy journals represents research activity. Even though we did not show this explicitly, we feel that this observation can be extended to refereed literature in general. The research cycle is a process common to many disciplines, and producing scholarly publications is always part of this cycle. Since all authors in astronomy are also users of the ADS and always use the ADS in their research, we feel that our results, presented in this paper, are of a more general nature and not just an indicator of trends in astronomy.

Is it meaningful to consider rankings of entities based on usage-based indicators? Making meaningful comparisons using citation-based indicators is already a complicated issue [18], so doing this using data that is intrinsically more noisy is going to be very hard indeed and probably even meaningless in a practical sense.

# Acknowledgements

This work has been supported by the NASA Astrophysics Data System project, funded by NASA grant NNX16AC86A.

# Author Information

Edwin A. Henneken

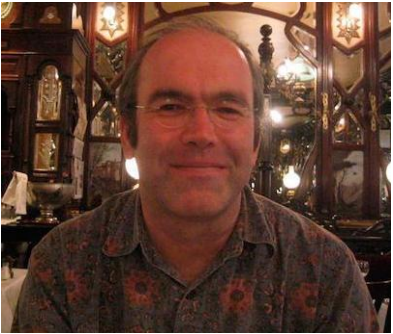

Address: Harvard-Smithsonian Center for Astrophysics, 60 Garden Street, Cambridge MA 02138, USA. E-mail: ehenneken@cfa.harvard.edu

Edwin Henneken received his scientific training at Leiden University (astrophysics) and the Vrije Universiteit Amsterdam (micro-meteorology), The Netherlands. After having worked in the Internet industry for a number of year, he joined in 2002 the Astrophysics Data System team, located at the Harvard-Smithsonian Center for Astrophysics in Cambridge, Massachusetts. In this role, he evaluates, integrates and enhances existing software systems used by the ADS project. He participates in curating the ADS holdings and managing the ADS digital preservation efforts. He uses the ADS holdings for informetrics and bibliometrics research. Currently, this work focuses on analysis of citation and usage data, literature recommendation and exposure of non-traditional products, like data and software.

Michael J. Kurtz

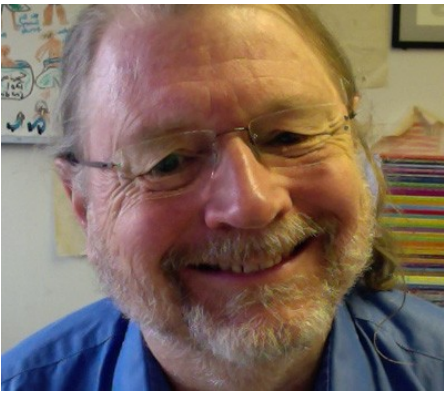

Address: Harvard-Smithsonian Center for Astrophysics, 60 Garden Street, Cambridge MA 02138, USA. E-mail: mkurtz@cfa.harvard.edu

Michael Kurtz is an astronomer and computer scientist at the Harvard-Smithsonian Center for Astrophysics in Cambridge, Massachusetts, which he joined after receiving a PhD in Physics from Dartmouth College in 1982.

Kurtz is the author or co-author of over 300 technical articles and abstracts on subjects ranging from cosmology and extra-galactic astronomy to data reduction and archiving techniques to information systems and text retrieval algorithms.

In 1988 Kurtz conceived what has now become the NASA Astrophysics Data System, the core of the digital library in astronomy, perhaps the most sophisticated discipline centered library extant. He has been associated with the project since that time, and was awarded the 2001 Van Biesbroeck Prize of the American Astronomical Society for his efforts.

He received the Citation Research Award from the American Society for Information Science and Technology, is an elected fellow in the Computer and Information Science section of the American Association for the Advancement of Science, and is an elected fellow in the Astrophysics division of the American Physical Society.

He is on the Board of Advisors, and is a founding member of Force11; is on the editorial board of the Journal of the Association for Information Science and Technology; and is the moderator for the Astrophysics Instruments and Methods section of arXiv.